\documentclass[preprint,aps,preprint,amsmath,amssymb]{revtex4-1}
\usepackage{graphicx}
\pdfoutput=1
\usepackage{amsmath}
\usepackage{color}
\usepackage{float}
\setcitestyle{super}
\usepackage[english]{babel}
\usepackage{color,soul}

\begin{document}
\title{Enhanced electrical transport through wrinkles in turbostratic graphene films}

\author{Monika Moun$^1$, Aastha Vasdev$^1$, Rajashekhar Pujar$^2$, K. Priya Madhuri$^3$, U. Mogera$^2$, Neena S. John$^3$}

\author{G.U. Kulkarni$^2$}
\author{Goutam Sheet$^1$}
\email{goutam@iisermohali.ac.in}

\affiliation{$^1$Department of Physical Sciences, Indian Institute of Science Education and Research (IISER) Mohali, Sector 81, S. A. S. Nagar, Manauli, PO: 140306, India}

\affiliation{$^2$Chemistry and Physics of Materials Unit and Thematic Unit on Nanochemistry, Jawaharlal Nehru Centre for Advanced Scientific Research, Jakkur P.O., Bangalore 560064, India}

\affiliation{$^3$Centre for Nano and Soft Matter Sciences, Jalahalli, Bengaluru 560013, India}

\begin{abstract}

\textbf{Formation of wrinkles is a common phenomenon in the large area growth of two dimensional layered materials on metallic substrates. Wrinkles can significantly affect the working of 2D materials based large scale electronic devices and therefore, it is of utmost importance to investigate local electrical properties of such wrinkled/folded structures on 2D materials. Here, we report local conductivity measurements by conducting atomic force microscopy (CAFM) and surface potential mapping by Kelvin Probe Force microscopy (KPFM) on large area wrinkled turbostratic graphene films grown on nickel foils. We show that the electrical transport current is several orders of magnitude higher on the wrinkles than that on the flat regions of the graphene films. Therefore, our results suggest that controlled engineering of such wrinkles on graphene may facilitate development of superior graphene-based nano-electronic devices where transport of high current through narrow channels are desired.}

\end{abstract}


\maketitle

Graphene, the first known 2D material has gained a huge amount of research attention in the recent past owing to its unique electrical, thermal, mechanical and optical properties.\cite{Castro, Balandin, Geim} It has shown its great potential in various electronic devices including field effect transistors\cite{Novoselov}, sensors \cite{Choi}, flexible devices \cite{Kim} and opto-electronic devices.\cite{Bablich} However, most of these have been achieved in small laboratory scale devices.  In order to manufacture such devices at industrial scale, it is most important to grow high quality graphene in large area and by cost effective techniques. Polycrystalline metallic films or foils are preferred as substrates for large area growth of graphene. For such substrates, randomly directed wrinkles form on graphene films. This is mainly due to the thermal coefficient difference between two contacting materials during cooling process of the synthesis protocol. The wrinkles thus formed are robust and they remain present even after the graphene films are transferred to other substrates through chemical/physical transfer routes\cite{Hattab, Chae}. In multilayer and single layer graphene films, the wrinkles naturally like to cause strain in the lattice and thus causes local lattice deformation.\cite{Ma, Pereira}. Several reports on few layers graphene considered the wrinkles as line defects and wrinkle-induced degradation of the electrical properties were observed.\cite{Clark, Schneider, Willke}. The degradation in the electrical performance has been attributed to electron-flexural phonon scattering and the inhomogeneous surface potential leading to potential barrier for charge carriers.\cite{Vasi, Ni} In single layer large area graphene, several reports have suggested better electrical transport along the wrinkles than that across them. Therefore, the wrinkles on graphene behave as 1D transport channels.\cite{Zhu, Orcid} It has also been reported that wrinkles in 2D layers can significantly tune the intrinsic properties of the material such as its band structure. Bandgap opening in graphene has been shown using Scanning Tunneling Spectroscopy with dI/dV spectra higher at wrinkled portion as compared to the flat regions of graphene.\cite{Lim} Gao \textit{et al.} also showed high conductivity on graphene wrinkles at charge neutrality point than that of monolayer part.\cite{Song} The earlier investigations on electrical characterization of wrinkled graphene were performed using two and four terminal measurements involving contact fabrication processes which might contribute to unwanted contamination to the samples thereby making the interprepation of transport data non-trivial.\cite{Zhu, Ni, Zhang} Therefore, it is most important to synthesize and characterize large area graphene films by clean methods free from chemical processing of the surface. Here, we use Conductive Atomic Force Microscopy (CAFM) and Kelvin Probe Force Microscopy (KPFM) to locally map the graphene samples directly grown on nickel foils. We detected wrinkles as well as flat regions on such graphene films. We show that the wrinkled areas have significantly enhanced conductivity than the flat areas of the graphene films. Our study can offer new perspectives towards electronic transport of large area graphene films for application in electronic devices.

\begin{figure}[h!]
	
	\includegraphics[width=0.5\textwidth]{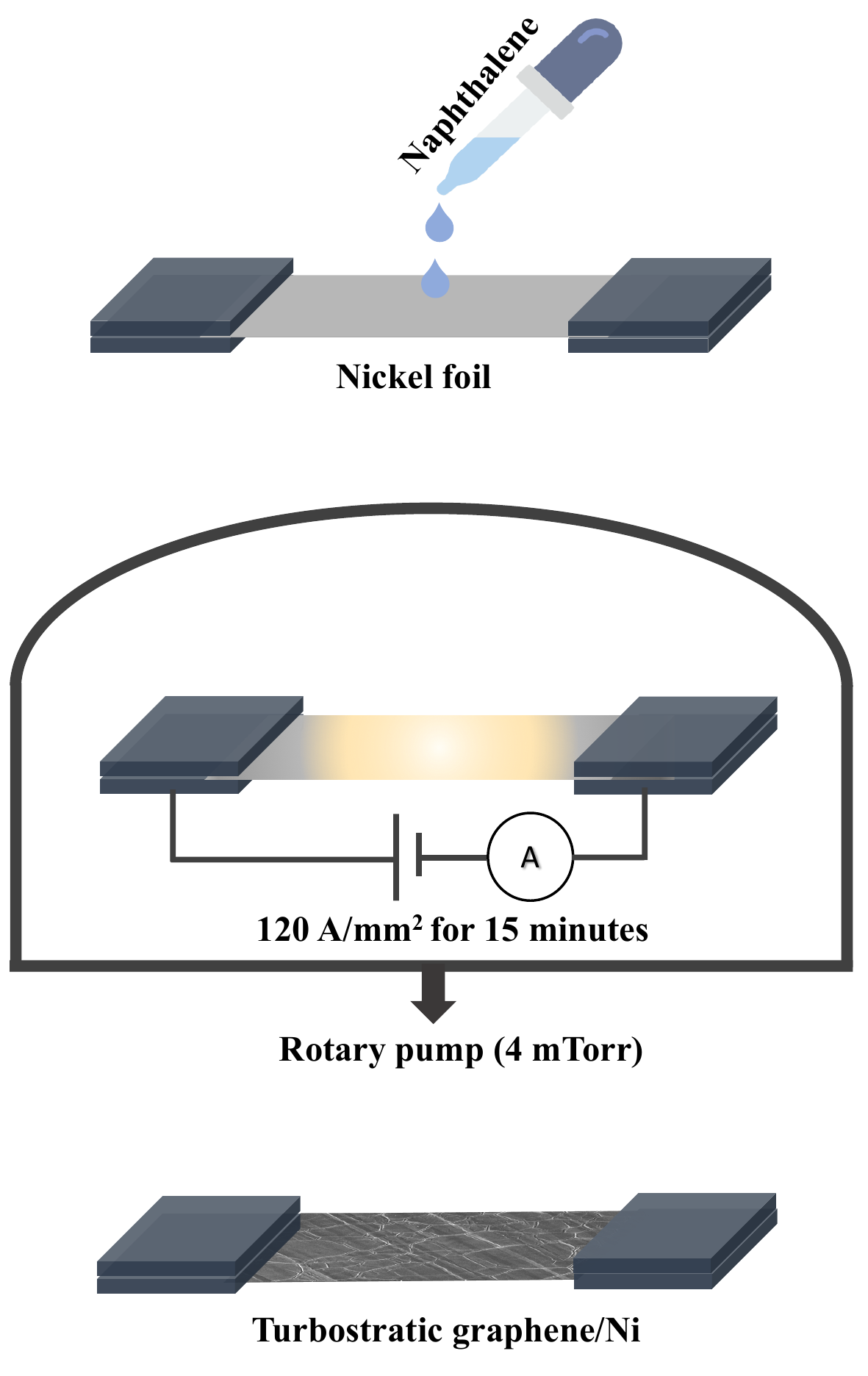}
	\caption{Schematic of the synthesis process of turbostratic graphene film on nickel foil.}
	\label{s1}
\end{figure}

For the synthesis of graphene film on nickel, a polycrystalline Ni foil was connected to the current-carrying electrodes and naphthalene solution in chloroform was drop cast onto the Ni foil. The Ni foil was Joule heated to red hot $(\sim800 ^{o}C)$ with a current density of 120 $A/mm^{2}$ using a DC source for 15 minutes at a pressure of 4 mTorr and then immediately cooled. Figure 1 shows the schematic of the synthesis process. The detailed synthesis process has been reported elsewhere.\cite{Kurra} In the as-synthesized graphene/nickel sample, a network of nano and microscopic wrinkles extending over centimetre scale on graphene film was observed along with flat graphene regions as shown in the SEM image (Figure 2a). As per the synthesis protocol used here, the as-synthesized multilayer graphene films are expected to behave as turbostratic graphene. Multilayer graphene films usually possess AB or ABC lattice stacking due to the inter-layer van der Waals interaction. Turbostratic graphene films consist of randomly oriented layers stacked with undefined order leading to decoupling of the stacked layers and increased interlayer distance. Hence in this context, it is considered as monolayer graphene. We investigated the turbostratic nature of the graphene films prior to our AFM experiments by Raman spectroscopy. The Raman spectra acquired on our samples exhibits signature of single layer graphene with high $I_{2D}/I_{G}$ ratio even though the sample is multilayer graphene. One such representative spectrum is shown in Figure 2(b). Decoupling of the graphene layers in such turbostratic graphene samples and the resemblance of the Raman spectra with single layer graphene was also reported earlier\cite{Mogera}. Figure 2(c-d) shows the cross-sectional SEM image of the graphene films transferred onto Poly-dimethylsiloxane (PDMS) substrate.

\begin{figure}[h!]
	
	\includegraphics[width=0.9\textwidth]{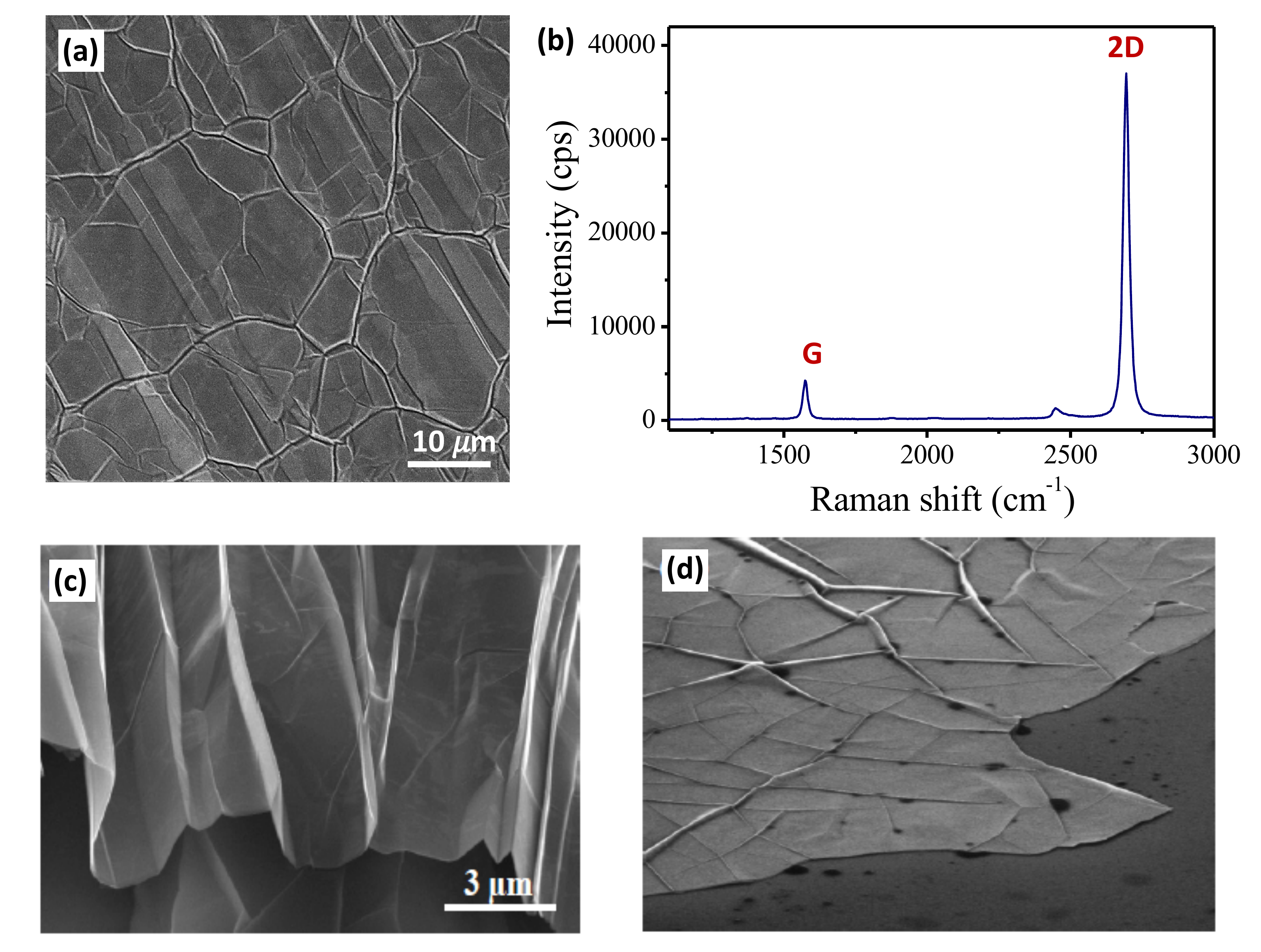}
	\caption{(a) Scanning electron microscopy of wrinkles present on graphene (b) Raman spectra of the sample (c-d) Cross sectional SEM images of graphene transferred on PDMS substrate.}
	\label{s1}
\end{figure}

Figure 3 shows the typical AFM image of the graphene layer clearly depicting the formation of wrinkles in the film. It is believed that the wrinkles form during the cooling process at the end of the high-temperature synthesis scheme as metal contracts more than graphene. The height of graphene wrinkles is in the range of 150-250 nm. The line profile in figure 3 shows that the wrinkles have higher thickness compared to flat graphene film.

\begin{figure}[h!]
	
	\includegraphics[width=0.9\textwidth]{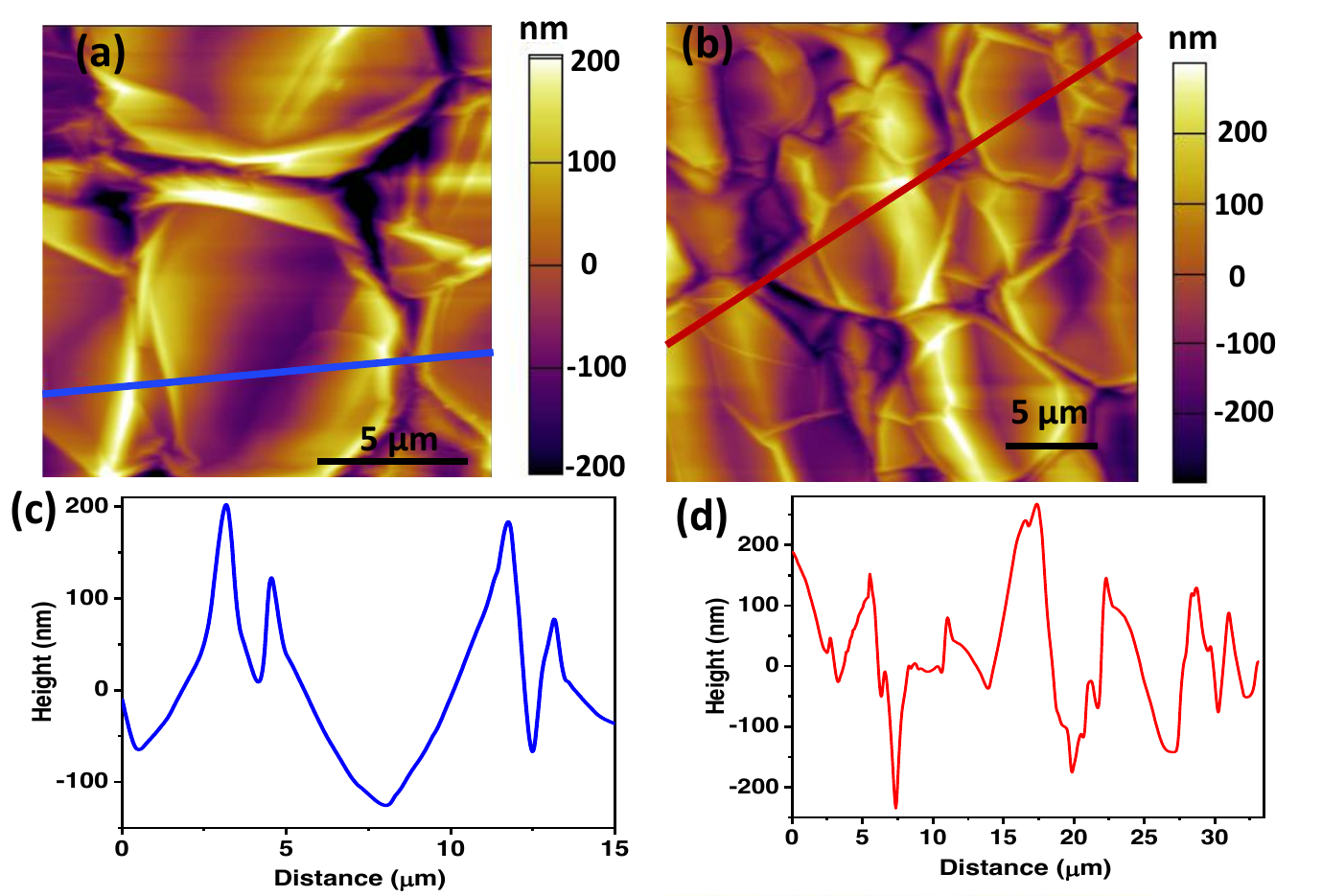}
	\caption{Morphology of graphene sheet with wrinkles and the corresponding line profile showing height variation in the as-synthesized graphene film.}
	\label{s1}
\end{figure}

Here, we try to investigate role of wrinkles on electronic properties of graphene sheet with the help of conducting atomic force microscopy (CAFM) technique. Figure 4 shows the morphology and corresponding current maps at different applied bias varying from 10 mV to 3V.  As the bias is applied, higher current is observed at the areas where graphene has wrinkles or it is folded. The current at the boundaries was found to increase gradually with the increasing applied bias as shown in the figure 4.
\begin{figure}[h!]
	
	\includegraphics[width=1.0\textwidth]{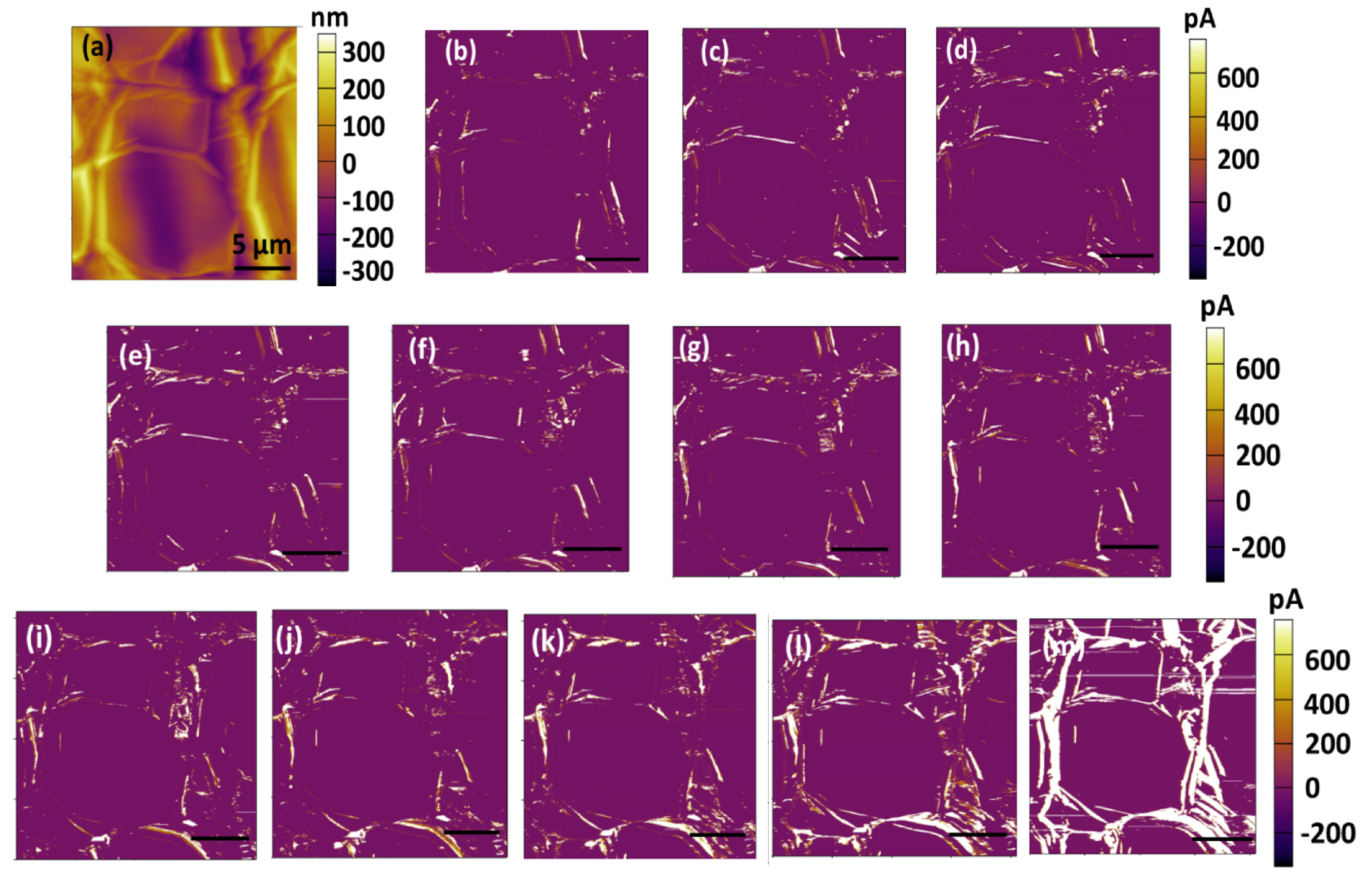}
	\caption{(a) Topography of graphene/nickel sample on 20 $\mu$m × 20 $\mu$m area. Conducting AFM images of the same area with varying applied bias (b) 10 mV (c) 20 mV (d) 30 mV (e) 40 mV (f) 50mV (g) 70 mV (h) 150 mV (i) 250 mV (j) 500 mV (k) 750 mV (l) 1V (m) 3V (scale bar : 5 $\mu$m).}
	\label{s1}
\end{figure}

As revealed in the AFM image (figure 3), the width and height of wrinkles vary from place to place. We observe that the wrinkles have higher height than the flat regions and are composed of folded graphene layers. At the wrinkles, multilayer stacking of graphene sheets can lead to lower resistance and increased conductivity by providing additional transport channels for charge carriers. Also, even on the same area, different wrinkles show variations in current values. Some wrinkles show comparatively higher current than the others. One possible cause for this variation may be the different thickness of the wrinkles. The multilayer/thicker wrinkles give higher current than the wrinkles with less number of layers. The folded features and curvatures induce lattice deformations and thus, the strain in the folded areas. The induced strain significantly modifies the electronic states and can lead to high conducting channels.\cite{Castro} Therefore, due to the multilayer additional transport channels and strain induced electronic states, the wrinkles can show high conductivity in two-dimensional graphene films and can be used for nano-electronic applications, such as interconnecting channels in chips and high speed graphene electronics.

Next, for better understanding of the charge transport, Kelvin potential force microscopy (KPFM) measurements were performed. KPFM is a technique used to detect the contact potential difference between a sharp AFM probe and the surface of the sample and thus the work function of the sample. It has been widely used to characterize monolayer, bilayer, and multilayer graphene due to substrate-induced work function variations.\cite{Yoo} KPFM measurements are performed in non-contact mode where the sample is grounded and bias (AC and DC) is applied to the AFM cantilever. The work function of the sample can be known by KPFM mapping by following equation\
\begin{center}
$\frac{\phi_{tip} -  \phi_{sample}}{e} = V_{CPD}$
\end{center}
where $\phi_{tip}$ is the work function of the tip, $\phi_{sample}$ is the work function of the sample and $V_{CPD}$ is the contact potential difference between tip and sample.\cite{Shen} Various factors such as wrinkles and grain boundaries in graphene are believed to significantly influence its local electrical properties. Figure 5 represents the surface potential mapping of wrinkled graphene sample. There was not observed any significant change in the surface potential and thus work function, as moving from flat graphene to wrinkles. In our case, work function of graphene film was calculated to be 4.85 eV. 
\begin{figure}[h!]
	
	\includegraphics[width=0.9\textwidth]{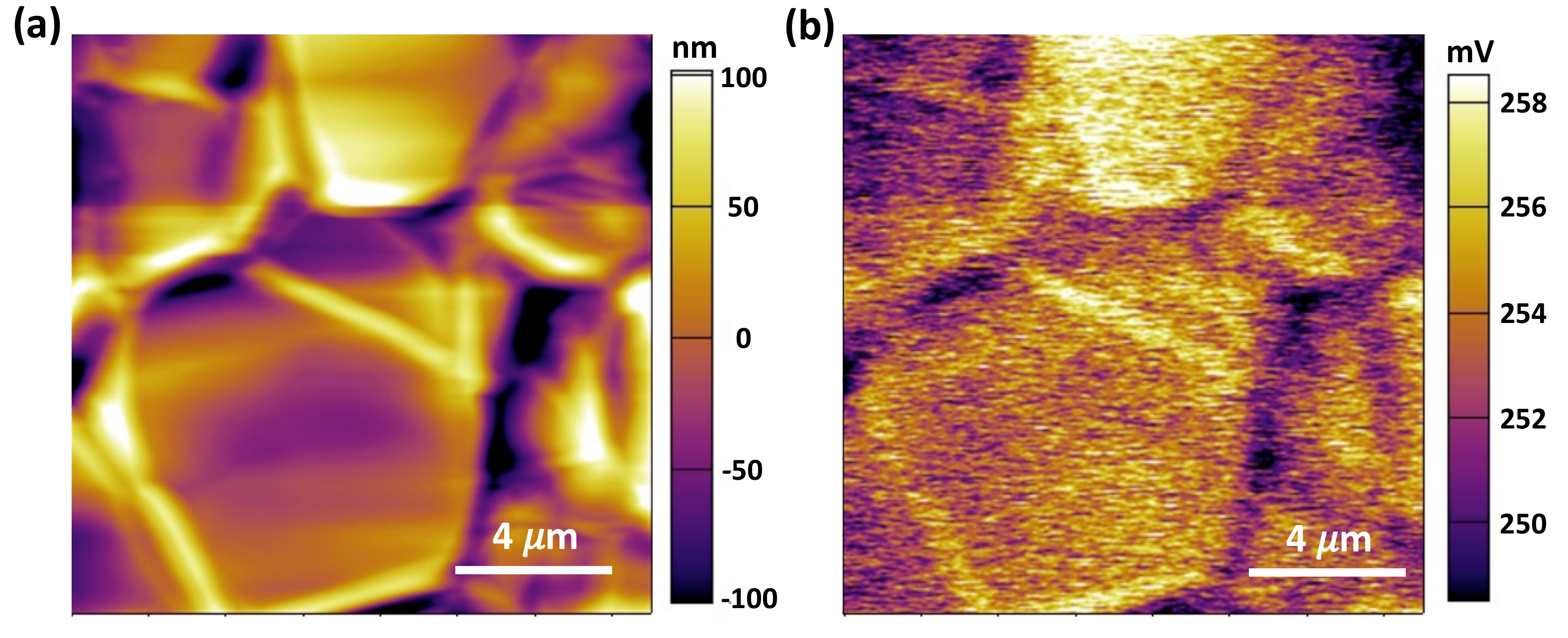}
	\caption {(a) AFM Topography of graphene/nickel sample on 15 $\mu$m × 15 $\mu$m area (b) corresponding KPFM mapping }
	\label{s1}
\end{figure}
From the surface potential mapping, one can deduce the change in the surface charge density if any. From figure 5, it is clear that there is no significant change in the surface potential as we go from flat region to the wrinkled region. Therefore, there is not noticable change in the surface charge density. But from the current map (figure 4), significant change in the current at the wrinkles can be clearly observed. The significant change in current can be attributed to either change in charge density or change in drift velocity. Change in charge density is not the case here, as seen from KPFM mapping. So, the change in current can be solely attributed to change in the drift velocity which suggests the change in the band structure of graphene at the wrinkled area. It has also been shown earlier that the graphene wrinkles can significantly alter the energy band structures and thus, the electrical properties of 2D graphene film.\cite{Deng, Chen}\\

In conclusion, we have shown that wrinkles, generally considered as quasi one-dimensional structures, can act as highly conducting channels in graphene films. KPFM measurements reveal less variation in surface potential from flat to wrinkled region and thus we can conclude that there is less change in charge density as one moves from the flat to the wrinkled areas. Hence, significant difference in current density at wrinkles can be attributed to change in band structure at wrinkles. Wrinkled graphene films offer a platform for 2D materials to broaden their applications in diverse fields. 

\section*{Methods}
Graphene films were chemically synthesized on polycrystalline nickel foils. Atomic force microscopy measurements along with its conducting mode and  Kelvin Probe Force Microscopy (KPFM) mode were performed using Atomic Force Microscope (MFP 3D of Asylum Research) in ambient conditions in contact mode and non-contact mode, respectively using Si cantilevers with Pt-Ir coating. The resonance frequency of AFM probes were around 75 kHz. Conducting Atomic force microscopy (CAFM) measurements were done in contact mode with bias applied to the sample. For KPFM experiments, firstly the AFM probe scanned the sample surface in tapping mode to obtain the topography. For surface potential mapping, bias was applied to the AFM cantilever. The AFM scanner lifted the probe $\sim$ 300 nm away from the surface, and scanned the sample surface in parallel fashion along with topography profile. For graphene/PDMS sample, the graphene layer was transferred from nickel to PDMS substrate using wet transfer method. The graphene film was separated from the Ni foil by overnight etching in aqueous $FeCl_{3}$ solution followed by repeated washing in water. 

\section*{Acknowledgements}
GS acknowledges financial support from Swarnajayanti fellowship awarded by the Department of Science and Technology (DST), Govt. of India (grant number DST/SJF/PSA01/2015-16). MM would like to acknowledge NPDF-SERB fellowship (PDF/2020/002122). AS would like to thank University Grants Commission (UGC), India for providing senior research fellowship.

\end{document}